\documentclass[twocolumn]{article}
\usepackage{graphicx}
\usepackage{amsmath}

\begin{document}

\title{The Tyranny of Qubits - Quantum Technology's Scalability Bottleneck}
\author{John Gough\\ jug@aber.ac.uk\\
Aberystwyth University, SY23 3BZ, Wales, United Kingdom}
\maketitle

\begin{abstract}
In this essay, I look at (bemoan) the issues surrounding simulating quantum systems in order to design quantum devices for quantum technologies. The program runs into a natural difficulty that simulating quantum systems really require a proper quantum simulator. The problem is likened to the \lq\lq tyranny of numbers\rq\rq that faced computer engineers in the 1960s. 
\end{abstract}

\section{Introduction}
\begin{quotation}
	\emph{Prediction is very difficult, especially about the future.}
	(Niels Bohr)
	\end{quotation}
Why don't we have a quantum computer by now? Quantum Computing has become the Holy Grail quest of Science and Technology in the 21st Century: it has attracted substantial attention from theoreticians and experimentalist; it has worked it way into the public consciousness; and its promises has influenced national governments. As a back up, the pursuit of quantum computing offers several spin-offs - quantum simulation, quantum enhanced sensing, quantum communications, etc. It has promised a step change in cyber-security, an exponential speed up in computation, and the solution to Big Data problems. Intensive and sustained lobbying has lead to several large scale programme investments by national and international funding agencies. 
But the idea is not new, and there have been experiments done for a long time now - so why do we not have even moderately sized quantum computers?

This special issue is concerned with \lq\lq quantum coherent feedback and reservoir engineering\rq\rq . In short, the use of feedback in quantum systems for technological goals. This is something I would consider to be a corner stone of quantum engineering (if for no other reason than that this is the way things pan out in the classical world of technology).
The field of quantum feedback is by now reasonably well-understood, both theoretically and mathematically, and its implementation is currently limited by two factors - 1) experimental realization, and 2) numerical simulation. There may be many answers that one could propose to the question in the opening, but I want to focus on the issue of these limitations as a quantum technology bottleneck.

\subsection{Experimental Directions}
The first of these, experimental realization, is one on which I cannot really give an expert opinion: there have be a small number of experiments showing how feedback can enhance performance of quantum systems (both for measurement-based quantum feedback \cite{WM_book,Sayrin}, and coherent quantum feedback \cite{KNPM10}-\cite{Crisafulli13}), but these seem to be hard to do and there has been no obvious programme to extend these results to quantum technology applications. My own experience from talking to experimentalists in Britain is that they dismiss feedback as impractical. This is somewhat  poignant given that Britain pioneered the use of feedback during the first great industrial revolution through the practical work of James Watt, and subsequently the theoretical work of James Clerk Maxwell\footnote{It has been suggested that the reason why feedback became so prominent in Britain during the first industrial revolution was in no small part due to its long tradition of empiricist philosophers such as Locke, Hobbes, Bacon and Hume \cite{Mayr}, and indeed due to the notion of self-regulation and equilibrium in society put forward by Hume and Adam Smith \cite{Richardson}: on continental Europe, in contrast, the prevailing philosophy was the Leibnizian one (as satirized by Voltaire in Candide) which favoured the notion of planning optimally and then fixing the plans. In other words, anticipating what we now understand as closed-loop (feedback) versus open-loop control. see also \cite{Hammond}.}.
Sadly, Britain does not seem set to use feedback in the quantum industrial revolution, though it is by no means alone in this position. It does not augur well that arguably the most important and useful concepts from classical engineering hardly features in the minds of most experimentalists in the emerging quantum technology sector.
Maybe this will change, but it ought to be a major worry nonetheless. 

To dwell on the issue of quantum feedback for a moment. There are two forms: measurement-based control which requires us to extract information from our system via measurements and actuate back on the system accordingly; and coherent quantum feedback control which involves coupling a suitably designed quantum controller to the system.
Intuitively, the latter is the more suggestive of an autonomous self-regulating system-controller pair similar to the Watt fly-ball governor. Unfortunately, it is the one with the least amount of experimental investigation.
We should mention the proposal of Kerckhoff \textit{et al}. \cite{KNPM10} who consider a qubit undergoing syndrome errors, but being regulated by a set of ancillary cavity modes - here everything is autonomous, there is no measurement, however, the syndrome errors are corrected by means of the specific coupling to the ancillary modes which is engineered so as to dissipate the appropriate qubit energy. Here the qubit and ancillary modes form a combined open system which is driven by coherent quantum input processes \cite{HP}-\cite{Gardiner}, and engineered to \lq\lq cool\rq\rq   the system back to the desired state.

My gut feeling is that quantum feedback and, more generally, the designability of quantum devices through quantum feedback network models must play a key role if the quantum world is ever to be an actual technology. This is at present a minority opinion - one that may turn out unfounded - based on the rather philosophical pointers above, but these principles do come with the weight of over 200 years of technological success behind them.

We can only hope that this type of control theoretic thinking eventually catches on in quantum technology. 

\subsection{Numerical Modelling}
The second factor, however, will be the main discussion point of this piece. The problem is a familiar one - quantum systems are notoriously difficult to simulate using a classical computer, so the problem of \textit{designing} quantum controlled systems becomes hard. As it stands, it is a bottleneck for Quantum Technology. To fully innovate quantum technologies, we need a complete theory of quantum control engineering that allows us to design engineered quantum devices, and for this we need to be able to simulated quantum systems and optimize over various realizations - if we cannot do this, then we do not have a future technology.

Retrospectively, the motivation for quantum computing is traced back to Richard Feynman's observation that simulation of quantum systems cannot be efficiently performed on classical computers, and his proposition that analogue quantum devices acting as universal quantum \textit{simulators} would lead to an exponential speed up compared to classical computers \cite{Feynman_QC}. The technological problem is that, in order to design a quantum 
computer, one first needs to be able to efficiently simulate classes of quantum systems that can only be efficiently simulated by a quantum computer!

Before thinking of ways around this, let's indulge in some diversion. A serviceable plot for a science fiction tale would be to have a time-traveler who uses a quantum computer to build a time machine to return to the 21st Century and, for whatever reasons, leaves his quantum computing device behind - the scientists (for some reason it's never the engineers in science fiction!) cannot reverse engineer the laptop but use its quantum computing capabilities to at least design a second functioning quantum computer - this inaugurates the quantum computing revolution eventually leading to our intrepid time 
traveler using the latest quantum computer to build a time machine, then going back in time and leaving his quantum laptop/mobile-phone/credit-card behind in the 21st Century.

But back now to the real world and we are left with the core question: \textit{how do we design a quantum computer without the aid of a quantum computer?}

\section{The Problem with Wiring Things Up}
\begin{quotation}
	\emph{For a successful technology, reality must take precedence over public relations, for Nature cannot be fooled.}
	(Richard Feynman)
\end{quotation}
It is not the first time that technology has faced a seemingly insurmountable hurdle. The more informed reader may have noticed 
that the title of this letter takes its queue from the phrase \emph{Tyranny of Numbers} coined by Jack Morton (vice President of Bell Labs) to describe the main problem facing the nascent computing industry: in Morton's own words (quoted from \cite{Love,Schwaderer}) ... \textit{For some time now, electronic man has known how \lq in principle\rq  to extend greatly his visual, tactile, and mental abilities through the digital transmission and processing of all kinds of information. However, all these functions suffer from what has been called \lq \textbf{the tyranny of numbers}\rq . Such systems, because of their complex digital nature, require hundreds, thousands, and sometimes tens of thousands of electron devices.} The problem has two aspects: a practical one that having workers try and solder connections onto transistor components was too labour intensive, costly and unreliable; and a pragmatic one that there was a level of complexity beyond which computers became too unreliable and error-prone to justify their computational advantage.

The essential problem here was that electronic circuits were becoming increasingly smaller scale and more reliable, but this resulted in increasingly more complex devices consisting on thousands of components (capacitors, resistors, transistors, etc.) with substantially more interconnection possibilities (which were themselves becoming more difficult to make). It was a massive effort to cut up the silicon blocks and attach these interconnections - often by hand. And later, as the number of components desired turned to millions, it soon became clear that the modelling problem itself was becoming intractable. Morton's own idea to tackle this
was through model and architectural reduction - that is, to for simplified ``functional devices'' possessing as few components and interconnections as possible while achieving as close to universal functionality as one reasonably can \cite{Gertner}.

As it turned out, this led nowhere. The solution to the problem - in this case making functional devices from silicon - was to fashion a complete circuit on chip: the integrated circuit! The idea had been around for a long time: an integrated circuit semiconductor amplifier had been patented by Werner Jacobi of Siemens as far back as 1949. Тhe concept of the integrated circuit had already been presented by radio engineer Geoffrey Dummer \cite{Schwaderer}: in 1952 he wrote \textit{I would like to take a peek into the future. With the advent of the transistor and the work on semi-conductors generally, it now seems possible to envisage electron equipment in a solid block with no connecting wires. The block may consist of layers of insulating, conducting, rectifying and amplifying materials, the electronic functions being connected directly by cutting out areas of the various layers}. It was only in 1958 that Jack Kilby, Robert Noyce and Jean Hoerni independently developed the first prototypes. 
(Kilby won the 2000 Nobel prize in Physics for the invention of the integrated circuit, though there were clearly major contributions from multiple researchers.)
The solution was remarkably simple: the patterned wafer \textit{is} the machine ... it doesn't need to be cut up into minute pieces only to be reassembled with lots of unreliable wires.

It is worth noting that the integrated chip itself was very slow to get off the ground \cite{Schwaderer}. 

\section{Quantum Interconnections and Scalability}
\begin{quotation}
	\emph{No, [of course, I don't believe the horseshoe brings good luck] ... but I am told it works even if you don't believe in it.}
	(Niels Bohr)
\end{quotation}
Quantum computers operate by means of qubits - quantum components with a 2-dimensional Hilbert space - and there are a variety of different ways to realize qubits physically. To be precise, these are the \textit{logical qubits}, and they are the staple concept of standard theories of quantum computation \cite{NC}. Quantum computation to a large extent deals with idealized models that are never found in Nature and this was quickly realized as being a potentially fatal obstacle to performing actual quantum computations. The operations - \textit{quantum gates} - performed on logical qubits are again an idealization not precisely available in Nature. Fortunately, Peter Shor \cite{Shor} was able to play one of the great get-out-of-jail-free cards with the proposal that quantum error correction is in principle possible. However, in order to achieve quantum error correction one has to perform a syndrome measurement to check whether the qubit has changed from its desired state - the syndrome measurement should to be done on auxiliary degrees of freedom (in Shor's scheme one needs 9 auxiliary qubits for each logical qubit) which are coupled to our logical qubit. If a syndrome error is detected, then an appropriate unitary gate is applied to restore the logical qubit back to the desired state. But we still have to engineer the auxiliary qubits, couple them to the logical qubits, perform potentially imprecise quantum gates - this is a limiting factor. The natural question is whether quantum error correction actually introduces more error than it corrects. The Quantum Fault Tolerance (or Quantum Threshold) Theorem of Aharonov and Ben-Or \cite{Threshold} shows that it is nevertheless possible to get a noisy quantum computer to simulate an idealized quantum computer to a given level of accuracy provided the error-rate is below a sufficient threshold level. We should mention that the applicability of the Quantum Threshold Theorem
to physical systems is not without criticism, notably by Robert Alicki on thermodynamical grounds \cite{Alicki_Hor} - \cite{Alicki13}.

The good news is that we can in principle run quantum algorithms on a noisy quantum computer with suitably engineered quantum error corrections: the bad news is that we'll need a lot more qubits to do this. The success of quantum computing, and indeed much of the lower hanging quantum technology fruit, therefore hinges on our ability to scale up quantum components. Unfortunately this is where things become even noisier, more process-intensive and hardware-intensive, and incoherent.

At present, there seems to be a limit of about 5 logical qubits for a quantum computer achieved so far, though with aspirations for hundreds, if not thousands, of qubits in the near future \cite{who}. It is clear that scalability is a concern amongst the quantum technology sector. There are several experimental proposals that claim to be scalable platforms. This may be true, and it may be that we will soon get our quantum computer if we are patient enough, but this is not my concern.

My contribution to quantum control has been primarily through the theory of quantum feedback networks developed with Matthew James
\cite{GouJam09a}-\cite{NM_QFN_I}. This is a modular theory of Markovian quantum input-output components that allows one to apply the block-design thinking from traditional engineering to quantum systems. For linear quantum components, one gets a bilinear control theory similar to the one found in linear systems theory, however, the theory is much more general ... the components can be qubits, spin systems, etc. The framework has been referred to as the SLH theory due to the fact that each component has its own internal Hamiltonian $H$, coupling (or collapse) operators $L$, and a unitary scattering matrix $S$ of quantum input to output fields. A computer package, QNET, has been developed by researchers at Stanford \cite{Tezak} to apply these rules for networks of standard components - qubits, cavity modes, beam-splitters, phase shifter, Kerr non-linearities, etc. Once the total network description has been determined, one may then apply simulation packages such as QuTiP.

On the one hand this leads us down a route that is highly suggestive of the standard approaches routinely used in classical technologies. The bad news, of course, is that we are dealing with quantum systems so simulation rapidly becomes a major issue. Even a moderately small number of simple quantum components in a quantum feedback network becomes impractical. Sadly, even Small-Scale-Integration design looks as infeasible in the quantum world. (For more discussion about these issues, see \cite{Santori}, \cite{Bowen2017}.) 

Where does this leave us? It could of course turn out to be a non-problem. It may happen that quantum technology goes on its present course and gets to the quantum supremacy stage without advanced quantum control theoretic thinking. I doubt this, but even if it does happen then we arrive at a situation where we have things working without knowing why exactly they are working or how best to improve them. Throughout the historical development of science and technology, this is a highly non-ideal situation. If we are expecting a new industrial revolution, then we need to be able to design the components otherwise we do not have a true technology. How we get to Medium, Large, or even Very-Large Quantum Scale Integration would be still anyone's guess.

\section{Outlook}
\begin{quotation}
	\emph{The search for the Grail is the search for the divine in all of us. But if you want facts, Indy, I've none to give you. At my age, I'm prepared to take a few things on faith.}
	(Marcus Brody: Indiana Jones and the Last Crusade)
\end{quotation}
An obvious way to try and get out of this conundrum is that we direct the attention of quantum simulation away from the proposed areas of investigation \cite{QSim} in the various quantum technology programmes, and instead use it as a resource for designing better quantum simulators. To the best of my knowledge, this has not been suggested up to now. 

It is not clear, however, whether this would work! The problem is somewhat akin to von Neumann's question of whether there existed universal constructors - machines capable of self-replication, or even building more complicated machines. There the answer is affirmative, but it needed the brilliance of von Neumann to show this. In contrast, the suggestion above is a practical one of using current state-of-the-art quantum machines as a design tool to improve the next generation. Like most things in quantum technology, this is uncharted territory.

On the other hand, the quantum feedback network theory may give some insight into the scalability issues - at least we may get some idea of figures of merit as to how noise, time delays, model imperfections, etc., scale up. Bounds on the performance as a function of number of components may reveal how drastic the scalability problem is for realistic situations.

Alternatively, it may happen that the solution is similar to that of the originally tyranny of numbers problem: we just start fabricating large integrated quantum circuits. The difficulty here is that the sub-components to be wired up ought to quantum dynamical systems - this is a hard thing to do experimentally. Most progress has been made on quantum circuits which act as static devices - effectively Mach-Zehnder networks consisting of up to about 20 beam-splitters - but an assembly of qubits on-chip communicating by quantum field signals still looks to be very far away.

There is general agreement that the scientists involved in quantum technology sector need to be cautious and realistic in their promises: this may be coming too late! One of the principle dangers of lobbying for science funding is that it creates a runaway positive feedback loop where success in attracting funding is proportional to the proposed impact. At any rate, the sector now has some pretty intimidating goals to make good. 

My view is that the central question here is whether quantum systems are \textit{technologizable} or not. By this, I mean whether the standard approaches of design and development that characterize modern industries can be applied to products that are quantum systems. Otherwise, we will be limited to a small number of applications lacking an overarching control engineering and product innovation framework. In principle, we now know that many of the core ideas of feedback control can be carried over into the quantum domain, but this has yet to progress beyond mathematical results, and a few proof-of-concept experiments. At the present time, one senses that there is view that a quantum computer is a collection of quantum gates, that being able to realize quantum gates means we are guaranteed to be able to build a quantum computer, and once that is done we can get engineers in to optimize performance. This is too naive! A good description of what a quantum systems engineering theory should look like has been given by Everitt \textit{et al.} \cite{Everitt}, but this still assumes a good enough understanding of quantum devices - for the time being we still do not know how to simulate even relatively low numbers of components in a quantum feedback network, let alone get to the device level.

It is worth looking at the birth of modern physics, in particular, quantum mechanics. Planck and Einstein were able to make their intellectual leaps because they were thinking like modern physicists. The reason why they were doing so was because they were following the ideas of Boltzmann, who is arguably the first to think like a modern physicist (see, for instance, \cite{everdell,ellis}), and who was the first to introduce probability as a staple into physics.
Modern physics happened because the right people were primed to think about problems in the right way. 
The same applies to industrial revolutions.

To close, it might be worth recalling how Bell Labs dealt with the problem of commercializing the transistor following the breakthrough experiments of Bardeen, Brattain and Shockley. In 1948 Jack Morton - the same who coined the phrase tyranny of numbers - was told by the head, who was about to go off on a one month tour of Europe, to come up with an action plan ready by the time he returned \cite{Gertner}. Morton's eventual report emphasized that innovation should be thought of as a total process ... \textit{ It is not just the discovery of new phenomena, nor the development of a new product or  manufacturing technique, nor the creation of a new market. Rather, the process is all these things acting together in an integrated way toward a common industrial goal} \cite{Gertner}. This subsequently became the blueprint for industrial innovation from that point onwards. Morton's vision was that the principle challenges for technology where reliability, reproducibility and designability of devices, \cite{Gertner} pg. 109.

Quantum systems may well be technologized, but it will require something beyond the current mainstream thinking. Areas like quantum feedback and control engineering indicate gaps that must be closed if this is to ever happen. The tyranny of qubits problem will likely remain a bottleneck on quantum technology roadmaps (whether sign-posted or not) for some time to come, however, by identifying the problem and some of the surrounding issues may be productive.

\section{Acknowledgements}
The author acknowledges various conversation, discussions and rants with colleagues on the subject of quantum technology over the years, however, accepts sole responsibility for the views expressed here.

\end{document}